\documentclass[12pt, twoside, openright]{article}
\usepackage{amsfonts,amsthm,amssymb,amsmath,amscd}
\usepackage{graphicx, float}
\usepackage{longtable}
\usepackage{ae}
\usepackage{fancybox}
\usepackage{multirow}
\usepackage{color}
\usepackage{slashbox}
\usepackage{multirow}
\usepackage{listings}

\usepackage[pdftex]{hyperref}
\usepackage{eurosym}
\newcommand{\bs}[1]{ \boldsymbol{#1} }
\usepackage{here}
\begin{document}

\newtheorem{theorem}{Theorem}[section]
\newtheorem{corollary}[theorem]{Corollary}
\newtheorem{lemma}[theorem]{Lemma}

\theoremstyle{definition}
\newtheorem{definition}[theorem]{Definition}
\newtheorem{folgerung}[theorem]{Corollary}
\newtheorem{rem}[theorem]{Remark}
\newtheorem{conclusion}[theorem]{Conclusion}
\newtheorem{example}[theorem]{Example}
\newtheorem{notation}[theorem]{Notation}
\newcommand{\Var}{\mbox{Var}}
\newcommand{\Cov}{\mbox{Cov}}
\newcommand{\RC}{\mbox{RC}}
\newenvironment{bew}{\begin{proof}[Proof]}{\end{proof}}

\title{Parameter uncertainty for integrated risk capital calculations based on normally distributed subrisks}
\author{Andreas Fr\"ohlich\footnote{Zentrales Aktuariat Komposit, R+V Allgemeine Versicherung AG, Raiffeisenplatz 1,
		65189 Wiesbaden, Germany, Email: \url{andreas.froehlich@ruv.de}} and Annegret Weng\footnote{Hochschule f\"ur Technik, 
		Schellingstr. 24
		70174 Stuttgart, Germany
		Tel.: 0049-(0)711- 8926-2730, Fax: 0049-(0)711-8926-2553
		Email: \url{annegret.weng@hft-stuttgart.de} (corresponding author)}}
\date{\today}
\maketitle
\begin{abstract}
In this contribution we consider the overall risk given as the sum of random subrisks $\bs X_j$ in the context of value-at-risk (VaR) based risk calculations. If we assume that the undertaking knows the parametric distribution family 
subrisk $\bs X_j=\bs X_j(\theta_j)$, but does not know the true parameter vectors $\theta_j$, the 
undertaking faces parameter uncertainty. 
To assess the appropriateness of methods to model parameter uncertainty for risk capital 
calculation we consider a criterion introduced in the recent literature. According to this criterion, 
we demonstrate that, in general, appropriateness of a risk capital model for each subrisk does 
not imply appropriateness of the model on the aggregate level of the overall risk.\\
For the case where the overall risk is given by the sum of normally distributed subrisks we prove a theoretical result leading to an appropriate integrated risk capital model taking parameter uncertainty into account. Based on the theorem we develop a method improving the approximation of the required confidence level simultaneously for both  - on the level of each subrisk as well as for the overall risk. 
\end{abstract}
\section{Introduction}\label{kap:1}
The overall risk of an undertaking can usually be viewed as the sum of its subrisks. If we interpret the subrisks as random variables $\bs X_j$, $1\le j\le m$, the overall risk is given by $\bs X_{sum}=\sum_{j=1}^m \bs X_j$.

\smallskip
In this contribution we assume that the undertaking knows the parametrized distribution family for each of the random variables $\bs X_j=\bs X_j(\theta_j)$, but can only estimate the unknown true parameter (vector) $\theta_j$ from historical data. In this case, the undertaking faces parameter uncertainty.

\smallskip
We focus on the effect of parameter uncertainty on value-at-risk based risk capital calculations. If we take the randomness of the historical data and, therefore, the randomness of the estimates $\hat{\bs \theta}_j$ of the true parameter vectors $\theta_j$ parametrizing the subrisks $\bs X_j$ into account, both the modelled standalone risk capital $\textbf{RC}_j=\RC(\hat{\bs \theta}_j)$ of each subrisk $\bs X_j$ as well as the overall risk capital itself are random variables. Their realizations depend on the historical data.\\
\smallskip
This leads to the following requirement compatible with the usual statutory regulations on value-at-risk based risk capital calculations (see, for example, article 101 of the Solvency EU framework directive):
\begin{definition}\label{def:definition1}
The value-at-risk based risk capital requirement \textbf{\RC} for a random loss $\bs X$ with respect to a given confidence level $\alpha$ should be modelled in such a way, that the random loss $\bs X$ does not exceed the risk capital requirement \textbf{\RC} with a probability of $\alpha$ - taking into account the randomness of both, the random variable $\bs X$ and the risk capital requirement $\textbf{\RC}$.\\
In this case the underlying risk capital model is called \textbf{appropriate to model $\bs X$ for the given confidence level $\alpha$}. 
\end{definition}
Definition \ref{def:definition1} has been formalized and appropriate methods modelling a risk capital requirement under parameter uncertainty have been proposed (see e.g. \cite{bignozzi, bignozzi2,froehlich2014,gerrard,tsanakas,pitera}).\\
For practical applications we search for a integrated risk capital model which is simultaneously appropriate in the sense of Definition \ref{def:definition1} to model each subrisk $\bs X_j$, $j=1,\ldots,m$, as well as the overall risk $\bs X_{sum}$. 

\smallskip
In this contribution we consider the parameter uncertainty in the case where the overall risk is given as the sum of its subrisks.  We derive two main results:
\begin{itemize}
\item  We demonstrate that appropriateness of the risk capital model on the level of each subrisk does not automatically yield appropriateness of the risk capital model on the level of the overall risk. A counterexample is already given by the simplest case, the bivariate normal distribution with independent subrisks (see Section \ref{kap:3}).
\item We then concentrate on the case of an overall risk $\bs X_{sum}=\sum \bs X_i$ where $(\bs X_1,\ldots,\bs X_m)$ is multivariate normally distributed. In this case we develop an integrated risk capital model which determines risk capitals simultaneously for both, the single subrisks and the overall risk, based on the joint distribution of the modelled subrisks, that meets the required solvency probabilities in good approximation (see Section \ref{sec:finv}).
\end{itemize}
Note that the multivariate normal distribution is still popular in practice, e.g. for modelling market risks in banking (cf. \cite{riskmetrics} or \cite{hull}, Subsection 13.2), for the reserve risk in non-life insurance \cite{england2006,gisler2006} or for risk factors for the non-financial risk for life insurances \cite{solvencyAssump}.\\
Our contribution shows that even the simple case of the multivariate normal distribution is not straight-forward. It seems natural to investigate first the simple case. Furthermore, we would like to draw attention to the problem of modelling parameter uncertainty for aggregate distributions motivating the investigation of more complex cases like non-negative and heavy-tailed distributions needed to model risks in the non-life insurance.


\section{Preliminaries}\label{kap:2}
We do not assume that the reader is familiar with \cite{froehlich2014} or \cite{gerrard} and we, therefore, recall the definition of an appropriate method for calculating the risk capital given in Section 2 in \cite{froehlich2014}.
\begin{notation}
In order to follow the arguments it is crucial to make a difference between random variables and their realizations. Throughout the paper all random variables are printed in \textbf{bold}.\\ 
Let $\bs X$ be a random variable describing the potential loss of the next business year. We assume that the undertaking knows the parametric distribution family $\{\bs X(\theta)| \theta\in I\subseteq\mathbb{R}^d\}$ of $\bs X$ but does not know  the true parameter $\theta_0\in I\subseteq \mathbb{R}^d$ with $\bs X=\bs X(\theta_0)$. For simplicity, we assume that $\bs X$ has an invertible cumulative distribution function denoted by $F_{\bs X}=F_{\bs X(\theta)}$. Consider the map $X:[0;1]\times I \rightarrow \mathbb{R}$ defined by $X(\xi,\theta):=F^{-1}_{\bs X(\theta)}(\xi)$ and note that we can write $X(\bs \xi,\theta):=F^{-1}_{\bs X(\theta)}(\bs \xi)$ to denote the random variable $\bs X(\theta)$ where $\bs \xi$ is an on $[0;1]$ uniformly distributed random variable.
\end{notation}
We further assume that the historical data is a sample drawn from a random vector $\bs D$  with known distribution function $F_{\bs D}=F_{\bs D(\theta)}$, but unknown parameter  $\theta$. We denote an observed realization of $\bs D$ by $D$. Typical examples in the existing literature (see \cite{bignozzi,froehlich2014,gerrard}) restrict to the special case $\bs D=(\bs X_1,\ldots,\bs X_n)$ where $\bs X_i\sim \bs X$ for $i=1,\ldots,n$ and $(\bs X_1,\ldots,\bs X_n)$ is independent of $\bs X$. \\
For a given set of data $D$ we assume that the undertaking models its required risk capital using a predictive distribution by a 2-step algorithm:
\begin{enumerate}
\item Determination of a distribution for the modelled parameter vector $\bs \theta_{sim}$: Using a suitable method $M$ and the observed historic sample $D$ determine a parameter distribution $\mathcal{P}=\mathcal{P}(D;M)$ for $\bs \theta_{sim}$ modelling the uncertainty about the unknown parameter $\theta$. 
\item Modelled risk: Set $\bs Y:=X(\bs \xi, \bs \theta_{sim})$ (cf. notation introduced of this section) where $\bs \xi$ is an uniformly distributed random variable on $[0;1]$ and independent of the modelled parameter $\bs \theta_{sim}$.
\end{enumerate}
Note that $\bs \theta_{sim}$ and hence $\bs Y$ depends on $M$ and $D$. We call $\bs Y=\bs Y(D;M)=X(\bs \xi, \bs \theta_{sim})$ the modelled risk and define the modelled risk capital requirement by
$$
\\RC(\alpha;D;M):=F_{\bs Y}^{-1}(\alpha).
$$

Note that the modelled risk capital $\textbf{\RC}:=\RC(\alpha; \bs D; M)$ is itself a random variable whose distribution depends on the distribution of the historical data. The formal interpretation of Definition \ref{def:definition1} is given as follows  (cf. \cite{froehlich2014, gerrard}):
\begin{definition}\label{def:zwei}
The method $M$ resp. the parameter distribution $\mathcal{P}$ are called \textbf{appropriate for the confidence level $\alpha$} if and only if
\begin{equation}\label{gl1}
P\left(\bs X\le \\RC(\alpha; \bs D;M)\right)= \alpha.
\end{equation}
The method $M$ resp. the probability distribution $\mathcal{P}$ are called \textbf{appropriate} if and only if they are appropriate for all $\alpha$ with $0<\alpha<1$. In this case we call the risk capital model using the method $M$ \textbf{appropriate to model $\bs X$}.\\
 An integrated risk capital model is called \textbf{appropriate} if it is appropriate to model both the subrisks $\bs X_j$, $j=1,\ldots,m,$ and the 
 overall risk $\bs X_{sum} = \sum \bs X_j$ are appropriate.
\end{definition}
In the next example we give a overview of known results on the appropriateness of several approaches to model parameter uncertainty for the case of a single risk $\bs X$. We assume that the historical data are i.i.d. realizations drawn from $\bs X$.
\begin{example}\label{ex:inv}
Consider a normally distributed risk $\bs X$ with mean 0 and fixed, but unknown standard deviation $\sigma$. Let us suppose that we observe $n$ independent realizations $x_1,\ldots,x_n$ of $\bs X$ and consider the unbiased estimate of the unknown parameter $\sigma^2$ given by
\begin{equation}\label{eq:sigma_hat1}
\hat{\sigma}^2=\frac{1}{n}\sum_{i=1}^n x_i^2.
\end{equation}
\begin{enumerate}
\item Suppose we neglect parameter uncertainty and set $\bs \sigma^{sim}\equiv \hat{\sigma}$. It can be shown that $\bs \sigma^{sim}\equiv \hat{\sigma}$ is not an appropriate parameter distribution in the sense of Definition \ref{def:zwei} (cf. \cite{gerrard}, Table 1). 
\item Since $x_i$ are realizations of random variables $\bs X_i\sim \bs X$, $\hat{\sigma}^2$ is a realization of a random variables $\hat{\bs \sigma}^2$. It is well-known that
\begin{equation}\label{eq:2}
\hat{\bs \sigma}^2=\frac{\sigma^2}{n}\cdot \bs C, \bs C\sim \chi^2(n)
\end{equation}
where $\chi^2(n)$ is the $\chi^2$-distribution with $n$ degrees of freedom.\\
This could be considered as the justification to model the parameter distribution $\mathcal{P}$ by the right hand side of Equation (\ref{eq:2}) (with $\sigma$ replaced by $\hat{\sigma}$).\\
However, this does not define an appropriate probability distribution in the sense of Definition \ref{def:zwei}. Table \ref{tab:boots} displays the solvency probabilities $P\left(\bs X\le \\RC(\alpha; \bs D;M)\right)$ for $n=10$, $\sigma=1$ and different confidence levels $\alpha$ determined using a Monte-Carlo simulation with 10,000 simulations of $\bs X$, 10,000 samples $D=(x_1,\ldots,x_n)$ and 10,000 simulations to determine $\\RC(\alpha;D;M)$ given the sample $D=(x_1,\ldots,x_n)$. Here, $\\RC(\alpha;D;M)$ is the $\alpha$-quantile of the random variable defined by $\hat{\bs \sigma}\cdot \bs Z^\prime$ with $\hat{\bs \sigma}$ defined by Equation (\ref{eq:2}) and independent of $\bs Z^\prime\sim N(0;1)$.
\begin{table}[H]
		\begin{center}
			\begin{tabular}{|c|c|c|c|c|}
				\hline
				confidence level $\alpha$ &90\% &95\% &99\% &99.5\%\\
				\hline
				\hline
				probability of solvency &88.09\% &93.51\% &98.33\% &99.22\%\\
				\hline
			\end{tabular}
		\end{center}
		\caption{Solvency probabilities for $n=10$, $\sigma=1$ and different confidence levels $\alpha$ for the parameter distribution given by (\ref{eq:2})}\label{tab:boots}
\end{table}

\item 
We apply the method proposed in \cite{froehlich2014} to determine a probability distribution of $\bs \sigma^{sim}$. In the sequel we call this approach ``inversion method''.\\
Let the sample $x_1,\ldots,x_n$ representing the observed data be fixed and let $\hat{\sigma}^2$ be the fixed estimate given by (\ref{eq:sigma_hat1}). We would like to determine a parameter distribution reflecting the uncertainty about the unknown parameter $\sigma$ given the estimate $\hat{\sigma}$. The inversion method suggests to invert Equation (\ref{eq:2}) and to set
\begin{align*}
\left(\bs \sigma^{sim}\right)^2=\hat{\sigma}^2\cdot \frac{n}{\bs C}
\end{align*}
with the observed $\hat{\sigma}$.\\
Let $\bs Y= \bs \sigma^{sim}\cdot \bs Z^\prime$, $\bs Z^\prime\sim N(0;1)$ independent of $\bs \sigma^{sim}$, be the modelled risk and let $F^{-1}_{\bs Y}(\alpha)$ be its $\alpha$-quantile. We set $\\RC(\alpha;\hat{\sigma};M):=F^{-1}_{\bs Y}(\alpha)$.\\
For the normal distribution it is shown in \cite{froehlich2014} that
$$
P\left(\bs X\le \\RC(\alpha;\bs D;M)\right)=\alpha
$$
for all $\alpha$ where the historical data $\bs D=(\bs X_1,\ldots,\bs X_n)$ is an i.i.d. random sample drawn from $\bs X$.\\
The method is based on the fiducial inference approach introduced by Fisher \cite{fisher1930}. For a discussion of the strengthens and weaknesses of the fiducial approach and new developments see \cite{zabell92}, \cite{hannig2016}.
\item Another possibility to construct an appropriate probability distribution in the sense of Definition \ref{def:zwei} is the Bayesian approach using the non-informative prior $\pi(\sigma)=\sigma^{-1}$. In \cite{gerrard} it is proven that the posterior distribution of $\bs \sigma$ defines an appropriate probability distribution. For the normal distribution this Bayesian posterior distribution coincides with the distribution of $\bs \sigma^{sim}$ obtained from the inversion method described above (see \cite{hora}).
\end{enumerate}
\end{example}
Note that the historical data $\bs D$ do not need to be an i.i.d. sample drawn from the random variable $\bs X$. In particular, in the case of the overall $\bs X_{sum}=\sum \bs X_j$ the historical data are usually used on the more granular level of the subrisks $\bs X_j$, $1\le j\le m$, as input for the risk capital model: Typically, one considers data $\bs D^{(j)}=(\bs X^{(j)}_1,\ldots,\bs X^{(j)}_{n(j)})$ where $\bs X^{(j)}_1,\ldots,\bs X^{(j)}_{n(j)}$ is an i.i.d. sample drawn from $\bs X^{(j)}$ for $j=1,\ldots,m$. In this case, Definition \ref{def:zwei} can be applied to the overall risk $\bs X_{sum}=\sum \bs X_j$ setting $\bs D=(\bs D^{(1)},\ldots, \bs D^{(m)})$.\

\section{Single subrisks versus the overall risk}\sectionmark{Single subrisks versus the overall risk}\label{kap:3}
The overall risk of an undertaking is usually determined by the aggregation of its single subrisks. In Section \ref{kap:2} it has already been pointed out, that in most situations the data are used to estimate the parameters on the granular level of the single subrisks. For an effective risk management we should not only consider the overall risk of an undertaking, but also assess the material subrisks.\\ 
In this section we demonstrate that an appropriate method for the single subrisks does not automatically yield an appropriate method on the level of the overall risk.\\
For illustrative purpose we restrict to the case of two independent, normally distributed random variables $\bs X_1$ and $\bs X_2$ with known mean equal to 0 and fixed, but unknown standard deviations $\sigma_1$ resp. $\sigma_2$. We assume that the undertaking observes historical data $(x_1^{(1)},\ldots,x_{n(1)}^{(1)})$ resp. $(x_1^{(2)},\ldots,x_{n(2)}^{(2)})$ with $n(1)=n(2)=n$ which are realizations of the independent samples $(\bs X_1^{(1)},\ldots,\bs X_{n}^{(1)})$ and $(\bs X_1^{(2)},\ldots,\bs X_{n}^{(2)})$ with $\bs X_i^{(j)}\sim \bs X_j$ for all $i$ and $j\in\{1,2\}$  and $\bs X_i^{(j)}$ independent of $\bs X_1$ and $\bs X_2$.\\
The undertaking would like to quantify the overall risk
$$
\bs X_{sum}=\bs X_1+\bs X_2
$$
using an appropriate risk capital model.\\
One way to determine an appropriate probability distribution for the unknown parameter is the Bayesian method. In \cite{gerrard}, the authors proved that the non-informative prior distribution $\pi(\sigma)=\sigma^{-1}$ yields an appropriate posterior distribution of the modelled parameter $\bs \sigma^{Bayes,sim}$ and an appropriate risk capital model with modelled risk $\bs Y:=X(\bs \xi;\bs \sigma^{Bayes,sim})$, $\bs \xi\in U(0;1)$ for a normally distributed random variable $\bs X$ if we use the maximum likelihood method for the parameter estimation of $\sigma$.\\
Let $\bs Y_j=X_j(\bs \xi_j,\bs \sigma_j^{Bayes,sim})$ with i.i.d. $\bs \xi_j\in U(0;1)$ for $j\in\{1,2\}$ and set $\\RC(\alpha;(x_1^{(j)},\ldots, x_{n}^{(j)});\mbox{Bayes}):=F_{\bs Y_j}^{-1}(\alpha)$, $j=1,2$. Appropriateness for the single subrisk model implies
$$
P\left(\bs X_j\le \\RC(\alpha;(\bs X_1^{(j)},\ldots, \bs X_{n}^{(j)});\mbox{Bayes})\right)=\alpha.
$$
However, defining the modelled overall risk by $\bs Y_{sum}:=\bs Y_1+\bs Y_2$ the risk capital $$\\RC(\alpha;\{(x_1^{(j)},\ldots, x_{n}^{(j)}):j=1,2\};\mbox{Bayes}):=F_{\bs Y_{sum}}^{-1}(\alpha)$$ does not yield an appropriate risk capital model in the sense of Definition \ref{def:zwei}: Table \ref{tab:2} displays the
probabilities of solvency
$$
P\left(\bs X_{sum}\le \\RC(\alpha;\{(\bs X_1^{(j)},\ldots, \bs X_{n}^{(j)}):j=1,2\};\mbox{Bayes}\right)
$$
determined experimentally using a Monte-Carlo simulation with 10,000 simulations of $\bs X_{sum}$, 10,000 samples $(x_1^{(j)},\ldots,x_n^{(j)})$ of size $n=10$ of each subrisk $\bs X_j$, $j\in\{1,2\}$ and 10,000 simulations of the modelled risk $\bs Y_{sum}$ to determine $\RC(\alpha;\{(x_1^{(j)},\ldots,x_n^{(j)}):j=1,2\};\mbox{Bayes})$.
\begin{center}
\begin{table}[H]
\begin{center}
\begin{tabular}{|c|c|c|c|c|}
\hline
confidence level &$\sigma_1$ &$\sigma_2$ &probability of solvency\\
\hline
\hline
\multirow{2}{*}{90\%}&1&1&91.07\%\\
&1&2&90.94\%\\
&1&10&90.50\%\\
\hline
\multirow{2}{*}{95\%}&1&1&95.74\%\\
&1&2&95.74\%\\
&1&10&95.08\%\\
\hline
\multirow{2}{*}{99\%}&1&1&99.41\%\\
&1&2&99.19\%\\
&1&10&99.05\%\\
\hline
\multirow{2}{*}{99.5\%}&1&1&99.82\%\\
&1&2&99.65\%\\
&1&10&99.59\%\\
\hline
\end{tabular}
\end{center}
\caption{Results for the probability of solvency $P\left(\bs X_{sum}\le \RC(\alpha;\{(\bs X_1^{(j)},\ldots, \bs X_{n}^{(j)}): j=1,2\};\mbox{Bayes}\right)$ using the Bayesian approach with non-informative prior to model the subrisks for $n=10$}\label{tab:2}
\end{table}
\end{center}
\begin{rem}\label{rem:3.1}
	Note that the inversion method proposed in \cite{froehlich2014} leads to the same unsatisfactory result since in this particular situation the Bayesian approach proposed in \cite{gerrard} coincides with the inversion method (cf.  \cite{hora}).

\end{rem}

\begin{conclusion}
Even in the most simple case where the overall risk is given as the sum of two independent, normally distributed subrisks an appropriate modelling of the subrisks in the sense of Definition \ref{def:zwei} does not ensure the appropriateness of the 
risk capital model for the overall risk.
\end{conclusion}
\section{Results for multivariate normal distributed random variables}\sectionmark{Multivariate normal distribution}  
\label{sec:finv}
In this section we present an adjustment of the inversion method proposed in \cite{froehlich2014} (see also Example \ref{ex:inv}) leading to an appropriate risk capital model for $\bs X_{sum}=\sum \bs X_i$ where $(\bs X_1,\ldots,\bs X_m)$ is multivariate normally distributed.\\
For the sake of clarity, we first concentrate on the case of independent random variables, but the results can be generalized to take correlation into account (see Remark \ref{rem:corr} below).\\
Let  $\bs X_j=\mu_j+\sigma_j\cdot \bs Z_j$, $1\le j\le m$, be independent, normally distributed random variables with unknown, but fixed parameters $(\mu_j,\sigma_j)$. We do not assume that the historical time series are all of the same length. Let $n(j)$ be the length of the observed sample $(x_1^{(j)},\ldots,x_{n(j)}^{(j)})$ drawn from $\bs X_j$.\\
Let 
$$
\hat{\mu}_j=\frac{1}{n(j)}\sum_{i=1}^{n(j)} x_i^{(j)} \mbox{ and } \hat{\sigma}_j^2=\frac{1}{{n(j)}-1}\sum_{i=1}^{n(j)} (x_i^{(j)}-\overline{x}^{(j)})^2.
$$ 
Hence,
\begin{equation}
\label{eq:inverse}
\hat{\bs \mu}_j=\mu_j+\frac{\sigma_j}{\sqrt{n(j)}}\cdot \bs \zeta_j\mbox{ and }\hat{\bs \sigma}^2=\sigma^2\cdot \bs M_j
\end{equation}
with independent random variables $\bs \zeta_j\sim N(0;1)$ and $\bs M_j\sim \frac{\chi^2(n(j)-1)}{n(j)-1}$, $1\le j\le m$.\\
Following the inversion method introduced in \cite{froehlich2014} by solving  Equations (\ref{eq:inverse}) for $(\mu_j,\sigma_j^2)$ and using independent modelled random variables $\bs\zeta_j^{\prime}\sim \bs \zeta_j$ and $\bs M_j^\prime\sim \bs M_j$, $1\le j\le m$,  we derive
$$
\bs \mu_j^{sim}=\hat{\mu}_j-\frac{\bs \sigma_j^{sim}}{\sqrt{n(j)}}\cdot \bs \zeta_j^\prime\mbox{ and }(\bs \sigma_j^{sim})^2=\frac{\hat{\sigma}_j^2}{\bs M_j^\prime}.
$$
Thus the modelled subrisks are defined by 
$$\bs Y_j=\bs \mu_j^{sim}+\bs \sigma_j^{sim}\cdot \bs Z_j^\prime,\quad 1\le j\le m$$
with i.i.d $\bs Z_j^\prime\sim N(0;1)$. All random variables $\bs Z_j$, $\bs \zeta_j$, $\bs M_j$, $\bs Z_j^\prime$, $\bs \zeta_j^\prime$ and $\bs M_j^\prime$ are independent of each other. Note that the independence of $\bs \zeta_j^\prime$ and $\bs M_j^\prime$ is motivated by the fact that the estimates $\hat{\bs \mu}_j$ and $\hat{\bs \sigma}_j$ are independent random variables (\cite{hogg}, p. 214-216).\\
To stress the dependency of the modelled subrisks on the data $(x_1^{(j)},\ldots, x_{n(j)}^{(j)})$ we also use the notation $\bs Y_j=\bs Y_j(x_1^{(j)},\ldots, x_{n(j)}^{(j)})$.\\

The inversion method leads to an appropriate risk capital model for the subrisks (cf. \cite{froehlich2014}). However, according to Remark \ref{rem:3.1} just defining $\bs Y_{sum}:=\sum_j \bs Y_j$ would not yield an appropriate risk capital model for the overall risk. For this reason we introduce the following correction factor: Set
\begin{equation}\label{eq:lambdahat}
\lambda_j=\frac{\sigma_j^2\cdot \frac{n(j)+1}{n(j)}}{\sum_{k=1}^m \sigma_k^2\cdot \frac{n(k)+1}{n(k)}}
\quad\mbox{ and }\quad
\hat{\lambda}_j=\frac{\hat{\sigma}_j^2\cdot \frac{n(j)+1}{n(j)}}{\sum_{k=1}^m \hat{\sigma}_k^2\cdot \frac{n(k)+1}{n(k)}}
\end{equation}
and define the stochastic correction factor
\begin{align*}
\bs a_{sim}&:=\left(\sum \frac{\hat{\lambda}_j}{\bs M_j^\prime}\cdot \sum \lambda_j\bs M_j^\prime\right)^{-\frac{1}{2}}\\
&=\left(\frac{\sum \hat{\sigma}_j^2\cdot \frac{n(j)+1}{n(j)}}{\sum (\bs \sigma_j^{sim})^2 \cdot \frac{n(j)+1}{n(j)} \cdot \sum\lambda_j\bs M_j^\prime}\right)^{\frac{1}{2}}.
\end{align*}
This choice of $\bs a_{sim}$ is motivated by the following theorem.
\begin{theorem}\label{thm:main}
Let $\bs X_{sum}=\sum \bs X_j$ and define the modelled risk
$\bs Y_{sum}^{mod}=\bs Y_{sum}^{mod}(\{(x_1^{(j)},\ldots, x_{n(j)}^{(j)}):1\le j\le m\})$ by
$$
\bs Y_{sum}^{mod}:=(1-\bs a_{sim})\cdot \sum \hat{\mu}_j+\bs a_{sim}\cdot \sum_j \bs Y_j(x_1^{(j)},\ldots, x_{n(j)}^{(j)}).
$$ 
Set $\RC\left(\alpha;\left\{\left(x_1^{(j)},\ldots, x_{n(j)}^{(j)}\right):1\le j\le m\right\};\mbox{mod}\right):=F^{-1}_{\bs Y_{sum}^{mod}}(\alpha).$\\
This defines an appropriate risk capital model taking parameter uncertainty into account, i.e.
$$
P\left(\bs X_{sum}\le \\RC\left(\alpha;\left\{\left(\bs X_1^{(j)},\ldots, \bs X_{n(j)}^{(j)}\right):1\le j\le m\right\};\mbox{mod}\right)\right)=\alpha.
$$
\end{theorem}
\begin{bew}
With independent random variables $\bs Z^\prime, \bs \zeta\sim N(0;1)$ we have
\begin{align*}
\bs Y_{sum}^{mod}&=(1-\bs a_{sim})\cdot \sum \hat{\mu}_j+\bs a_{sim}\sum\bs Y_j\\
&\sim \sum \hat{\mu}_j+\bs a_{sim}\cdot \left(\sum -\frac{\bs \sigma_j^{sim}}{\sqrt{n(j)}}\cdot \bs \zeta_j^\prime+\bs \sigma_j^{sim} \cdot \bs Z_j^\prime\right)\\
&\sim \sum \hat{\mu}_j+\bs a_{sim}\cdot \sqrt{\sum (\bs \sigma_j^{sim})^2+\sum \frac{(\bs \sigma_j^{sim})^2}{n(j)}}\cdot \bs Z^\prime\\
&=\sum \hat{\mu}_j+\sqrt{\frac{\sum \hat{\sigma}_j^2\cdot \frac{n(j)+1}{n(j)}}{\sum (\bs \sigma_j^{sim})^2\frac{n(j)+1}{n(j)}\cdot \sum \lambda_j\bs M_j^\prime}}\cdot \sqrt{\sum (\bs \sigma_j^{sim})^2\cdot \frac{n(j)+1}{n(j)}}\cdot \bs Z^\prime\\
&=\sum \hat{\mu}_j+\sqrt{\frac{\sum \sigma_j^2\cdot \frac{n(j)+1}{n(j)}\cdot \sum \hat{\sigma}_j^2\cdot \frac{n(j)+1}{n(j)}}{\sum \sigma_j^2\cdot \frac{n(j)+1}{n(j)}\cdot \bs M_j^\prime}}\cdot \bs Z^\prime\\
&= \sum \mu_j+\sqrt{\sum \frac{\sigma_j^2}{n(j)}}\cdot \zeta+\sqrt{\frac{\sum \sigma_j^2\cdot \frac{n(j)+1}{n(j)}\cdot \sum \sigma_j^2\cdot \frac{n(j)+1}{n(j)}\cdot M_j}{\sum \sigma_j^2\cdot \frac{n(j)+1}{n(j)}\cdot \bs M_j^\prime}}\cdot \bs Z^\prime
\end{align*}
where $M_j:=\hat{\sigma}_j^2/\sigma_j^2$ is a realization of a $\chi^2(n(j)-1)/(n(j)-1)$ distributed random variable $\bs M_j$ and $\zeta:=\sum (\hat{\mu}_j-\mu_j)/\sqrt{\sum \sigma_j^2/n(j)}$ is a realization of a standard normally distributed random variable $\bs \zeta$.
Let $\sigma_{sum}^2=\sum \sigma_j^2$ and set $G(\{(x_1^{(j)},\ldots, x_{n(j)}^{(j)}):1\le j\le m\})=F_{\bs Y_{sum}^{mod}(\{(x_1^{(j)},\ldots, x_{n(j)}^{(j)}):1\le j\le m\})}(\bs X_{sum})$. We consider the random variable $G\left(\left\{\left(\bs X_1^{(j)},\ldots, \bs X_{n(j)}^{(j)}\right):1\le j\le m\right\}\right)$. With some algebraic manipulations using properties of the normal distribution it follows that
\begin{align*}
&G\left(\left\{\left(\bs X_1^{(j)},\ldots, \bs X_{n(j)}^{(j)}\right):1\le j\le m\right\}\right)\\
&= F_{\sum \mu_j+\sqrt{\sum \frac{\sigma_j^2}{n(j)}}\cdot \bs \zeta+\sqrt{\frac{\sum \sigma_j^2\cdot \frac{n(j)+1}{n(j)}\cdot \sum \sigma_j^2\cdot \frac{n(j)+1}{n(j)}\cdot \bs M_j}{\sum \sigma_j^2\cdot \frac{n(j)+1}{n(j)}\cdot \bs M_j^\prime}}\cdot \bs Z^\prime}\left(\sum \mu_j+\sigma_{sum}\bs Z\right)\\
&\sim F_{\sqrt{\frac{\sum \sigma_j^2\cdot \frac{n(j)+1}{n(j)}}{\sum \sigma_j^2\cdot \frac{n(j)+1}{n(j)}\cdot \bs M_j^\prime}}\cdot \bs Z^\prime}\left(\frac{-\sqrt{\sum \frac{\sigma_j^2}{n(j)}}\cdot \bs \zeta+\sqrt{\sum \sigma_j^2}\cdot \bs Z}{\sqrt{\sum \sigma_j^2\cdot \frac{n(j)+1}{n(j)}\cdot \bs M_j}}\right)\\
&\sim F_{\sqrt{\frac{\sum \sigma_j^2\cdot \frac{n(j)+1}{n(j)}}{\sum \sigma_j^2\cdot \frac{n(j)+1}{n(j)}\cdot \bs M_j^\prime}}\cdot \bs Z^\prime}\left(\sqrt{\frac{\sum \sigma_j^2\cdot \frac{n(j)+1}{n(j)}}{\sum \sigma_j^2\cdot \frac{n(j)+1}{n(j)}\cdot \bs M_j}}\cdot \bs Z\right).
\end{align*}
It follows that $G\left(\left\{\left(\bs X_1^{(j)},\ldots, \bs X_{n(j)}^{(j)}\right):1\le j\le m\right\}\right)$ is uniformly distributed on $[0;1]$. Hence,
\begin{align*}
&P\left(\bs X_{sum}\le \RC\left(\alpha;\left\{\left(\bs X_1^{(j)},\ldots, \bs X_{n(j)}^{(j)}\right):1\le j\le m\right\};\mbox{mod}\right)\right)\\
&=P\left(G\left(\left\{\left(\bs X_1^{(j)},\ldots, \bs X_{n(j)}^{(j)}\right):1\le j\le m\right\}\right)\le \alpha\right)=\alpha.
\end{align*}
\end{bew}
\begin{rem}
\begin{enumerate}
\item Note that the method works analogously for the maximum likelihood estimate of $\sigma^2$ by taking $\bs M_j\sim \frac{\chi^2(n(j)-1)}{n(j)}$.
\item Note that $\bs \sigma_j^{mod}:=\bs a_{sim}\cdot \bs \sigma_j^{sim}$ and $\bs \mu_j^{mod}:=\hat{\mu}_j+\frac{\bs \sigma_j^{mod}}{\sqrt{n(j)}}\cdot \bs \zeta_j^\prime$ defines a parameter distribution according to the modelled risk $\bs Y_j^{mod}=(1-\bs a_{sim})\cdot \hat{\mu}_j+\bs a_{sim}\cdot \bs Y_j$. Due to the multiplication by $\bs a_{sim}$ the modelled simulated standard deviations $\bs \sigma_1^{mod}$ and $\bs \sigma_2^{mod}$ are not uncorrelated, but the correlation is negligible for practical applications. Despite this correlation the random variables $\bs Y_j$ are still uncorrelated.
\end{enumerate}
\end{rem}
\begin{rem}
In practice, the weights $\lambda_j$ and hence the adjustment $\bs a_{sim}$ are unknown. Therefore, we use the estimate $\hat{\lambda}_j$ of $\lambda_j$ (cf. Equation (\ref{eq:lambdahat})). We denote the approximation of $\bs a_{sim}$ with $\hat{\lambda}_j$ instead of $\lambda_j$ by $\hat{\bs a}_{sim}$. Let $\hat{\bs Y}_{sum}^{mod}$ be the corresponding modelled risk and set $\\RC(\alpha;\{(x_1^{(j)},\ldots,x_{n(j)}^{(j)}): 1\le j\le m\};mod):=F_{\hat{\bs Y}_{sum}^{mod}}^{-1}(\alpha)$.\\
For $m=2$, Table \ref{tab:results} on p. \pageref{tablalognw} displays the probability 
$$
P(\bs X_{sum}\le\RC(\alpha;\{(\bs X_1^{(j)},\ldots,\bs X_{n(j)}^{(j)}): 1\le j\le m\};mod))
$$
for 10,000 simulations of $\bs X_{sum}$ and of the samples of size $n(j)=n_j$ of each subrisk $\bs X_i^{(j)}$, $1\le j\le m$, and 10,000 simulations of the modelled risk to determine the $\alpha$-quantile of the mixed random variable $\hat{\bs Y}_{sum}^{mod}$.
\begin{table}
\begin{center}
\begin{tabular}[t]{|c||c|c|c||c|}
\hline
\multirow{3}{*}{\backslashbox{$n_1,n_2$}{Parameter}} &\multirow{3}{*}{ $\sigma_1$} &\multirow{3}{*}{ $\sigma_2$}  &\multirow{3}{*}{$\alpha$} &probability of solvency\\
&&&&modelling (not modelling)\\
&&&&parameter uncertainty\\
\hline
\hline
\multirow{12}{*}{$n_1=n_2=10$}  &	1	&	0,1	&\multirow{3}{*}{$90,00\%$}	&	90,01\% (87,41\%)\\
&	1	&	1	&	&89,97\% (88,12\%)\\
&	1	&	2	&	&90,01\% (87,88\%)\\
\cline{2-5}
&	1	&	0,1	&	\multirow{3}{*}{$95,00\%$}		&94,96\% (92,49\%)\\
&	1	&	1	&	&95,06\% (93,28\%)\\
&	1	&	2	&	&94,96\% (93,02\%)\\
\cline{2-5}
&	1	&	0,1	&\multirow{3}{*}{$99,00\%$}		&99,00\% (97,36\%)\\
&	1	&	1	&	&99,07\% (98,02\%)\\
&	1	&	2	&	&99,01\% (97,81\%)\\
\cline{2-5}
&	1	&	0,1	&\multirow{3}{*}{$99,50\%$}		&99,50\% (98,22\%)\\
&	1	&	1	&	&99,54\% (98,78\%)\\
&	1	&	2	&	&99,49\% (98,61\%)\\ 
\hline
 \hline
\multirow{12}{*}{$2\cdot n_1=n_2=10$} &	1	&	0,1	&\multirow{3}{*}{$90,00\%$}		&	89,97\% (84,75\%)\\
&	1	&	1	&	&90,08\% (87,18\%)\\
&	1	&	2	&	&90,08\% (87,57\%)\\
\cline{2-5}
&	1	&	0,1	&\multirow{3}{*}{$95,00\%$}		&94,92\% (89,73\%)\\
&	1	&	1	&		&95,07\% (92,38\%)\\
&	1	&	2	&		&95,04\% (92,74\%)\\
\cline{2-5}
&	1	&	0,1	&\multirow{3}{*}{$99,00\%$}		&98,79\% (95,07\%)\\
&	1	&	1	&		&99,05\% (97,42\%)\\
&	1	&	2	&	&99,05\% (97,62\%)\\
\cline{2-5}
&	1	&	0,1	&\multirow{3}{*}{$99,50\%$}	&99,31\% (96,19\%)\\
&	1	&	1	&	&99,53\% (98,30\%)\\
&	1	&	2	&	&99,54\% (98,46\%)\\
\hline
\hline
\multirow{12}{*}{$2\cdot n_1=n_2=20$}&	1	&	0,1	&\multirow{3}{*}{$90,00\%$}	&90,02\% (87,40\%)\\
&	1	&	1	&	&90,03\% (88,59\%)\\
&	1	&	2	&&90,09\% (88,81\%)\\
\cline{2-5}
&	1	&	0,1	&\multirow{3}{*}{$95,00\%$}	&95,01\% (92,48\%)\\
&	1	&	1	&	&95,05\% (93,74\%)\\
&	1	&	2	&&95,14\% (93,91\%)\\
\cline{2-5}
&	1	&	0,1	&\multirow{3}{*}{$99,00\%$}	&99,01\% (97,35\%)\\
&	1	&	1	&	 &99,02\% (98,31\%)\\
&	1	&	2	&	 &99,04\% (98,38\%)\\
\cline{2-5}
&	1	&	0,1	&\multirow{3}{*}{$99,50\%$}	&99,50\% (98,21\%)\\
&	1	&	1	&	&99,50\% (99,00\%)\\
&	1	&	2	&		&99,51\% (99,05\%)\\
\hline
\end{tabular}
\caption{$P(\bs X_{sum}<F_{\hat{\bs Y}_{sum}^{mod}}^{-1}(\alpha))$ for different values of $\mu_1$, $\mu_2$, $\sigma_1$, $\sigma_2$, $n_1$, $n_2$ and $\alpha$.}\label{tab:results}
\label{tablalognw}
\end{center}
\end{table}
\end{rem}
\begin{rem}\label{rem:corr}
The assertion of Theorem \ref{thm:main} can be generalized for the multivariate normal distributed with correlated subrisks with known correlation $\rho_{ij}$.\\
We assume that the observed data exhibits the same correlation $\rho_{ij}$ and but that the random variables $\bs X_{i}^j$ and $\bs X_l^{k}$ are uncorrelated for different points in time.\\
We then use the same correlation $\rho_{ij}$ for the random variable $\bs Z_j^\prime$  and the adjusted correlation $\rho_{ij}\cdot \frac{\min(n(i),n(j)}{\sqrt{n(i)n(j)}}$ for the random variables $\bs \zeta_j^\prime$ to take into account the length of the different time series. The correlation does not effect the random variables $\bs M_j^\prime$.\\
The procedure is than analogous to the case with independent subrisks. We only need to adjust the correction factor $\bs a_{sim}$: Define the generalized weights
$$
\lambda_{ij}=\frac{\rho_{ij}\sigma_i\sigma_j\cdot \left(1+\frac{\min(n(i),n(j))}{n(i)n(j)}\right)}{\sum_{k,l=1}^m \rho_{kl}\sigma_k\sigma_l\cdot \left(1+\frac{\min(n(k),n(l))}{n(k)n(l)}\right)}
$$
and
$$
\hat{\lambda_{ij}}=\frac{\rho_{ij}\hat{\sigma_i}\hat{\sigma_j}\cdot \left(1+\frac{\min(n(i),n(j))}{n(i)n(j)}\right)}{\sum_{k,l=1}^m \rho_{kl}\hat{\sigma_k}\hat{\sigma_l}\cdot \left(1+\frac{\min(n(k),n(l))}{n(k)n(l)}\right)}
$$
and set
\begin{align*}
\bs a_{sim}&:=\left(\sum_{i,j=1}^m \frac{\hat{\lambda}_{ij}}{\sqrt{\bs M_i^\prime\bs M_j^\prime}}
\cdot \sum_{i,j=1}^m\lambda_{i,j}\sqrt{\bs M_i^\prime\bs M_j^\prime}
\right)^{-\frac{1}{2}}\\
&=\left(\frac{\sum_{i,j=1}^m\rho_{ij}\hat{\sigma_i}\hat{\sigma_j}\cdot \left(1+\frac{\min(n(i),n(j))}{n(i)n(j)}\right)}{\sum_{i,j=1}^m\rho_{ij}\bs \sigma_i^{sim}\bs\sigma_j^{sim}\cdot \left(1+\frac{\min(n(i),n(j))}{n(i)n(j)}\right) \cdot \sum_{i,j=1}^m \lambda_{ij}\sqrt{\bs M_i^\prime\bs M_j^\prime}}\right)^{\frac{1}{2}}.
\end{align*}
\end{rem}
We summarize the results of this subsection:\\
The algorithm described above consists of three steps:
\begin{enumerate}
	\item Use the inversion method to calculate the risk capitals $F_{\bs Y_i}^{-1}(\alpha)$, $i=1,\ldots,m,$ for the subrisks. 
	\item Determine the pathwise value of $\hat{\bs a}_{sim}$.
	\item Derive the pathwise realization of the aggregate variable $\hat{\bs Y}_{sum}^{mod}=(1-\hat{\bs a}_{sim})\sum \hat{\mu}_j+\hat{\bs a}_{sim}\sum \bs Y_j$ and calculate the overall risk capital
	$F_{\hat{\bs Y}_{sum}^{mod}}^{-1}(\alpha)$.
\end{enumerate}
Note that:
\begin{enumerate}
\item The modelled subrisks $\bs Y_i$ lead to an appropriate risk capital model in the sense of Definition \ref{def:zwei} for the single subrisks $\bs X_i$. Hence, our approach allows to appropriately evaluate the risk capital for the subrisks. 
\item Using the adjustment with the stochastic correction factor $\bs a_{sim}$ in Theorem \ref{thm:main} we get an appropriate risk capital model according to Definition \ref{def:zwei} for the overall risk. Thus, the integrated risk capital model is appropriate in the sense of Definition \ref{def:zwei}. However, the true factor $\bs a_{sim}$ requires the knowledge of the true parameters $\sigma_j$ resp. $\lambda_j$. 
\item On the level of the overall risk, the approximation using $\hat{\bs a}_{sim}$ instead of $\bs a_{sim}$ cannot be proven to be (exactly) appropriate in the strict sense of Definition \ref{def:zwei}. However, the experimental results in Table \ref{tab:results} show that the required confidence level is generally achieved in good approximation. Only in the case where the sample size is very small and the ratio of the standard deviations differs significantly from $1$ (c.f. $2n_1=n_2=10$, $\sigma_1=1=10\cdot \sigma_2$ and $\alpha=99.5\%$) the probability of solvency is significantly lower than the required confidence level. But even for this exceptional case the results are much better than without modelling parameter uncertainty.
\end{enumerate}

\section{Conclusion}
This contribution deals with parameter uncertainty in the context of integrated value-at-risk based risk capital calculations. We give evidence that appropriateness of the risk capital model for each subrisk in the sense of Definition \ref{def:zwei} does not imply the appropriateness of the overall risk capital model.\\
Then, we presented a new method to model the risk capital requirement in the case where the overall risk is given by $\bs X_{sum}=\sum \bs X_i$ where $(\bs X_1,\ldots,\bs X_m)$ is multivariate normally distributed. In Theorem \ref{thm:main} we prove that it yields an appropriate risk capital model for the single subrisks as well as the overall risk based on the modelled distribution of the subrisks. For this purpose we had to introduce the stochastic correction factor $\bs a_{sim}$ which requires the knowledge of the unknown parameters and is, therefore, replaced by the estimation $\hat{\bs a}_{sim}$. In Table \ref{tab:results} we present experimental results using the approximation $\hat{\bs a}_{sim}$.\\
Our article takes a first step towards finding a risk capital model taking parameter uncertainty into account, which attains the required probability of solvency simultaneously for both the overall risk as well as for all subrisks. We hope that it encourages future research in this direction.However, it is still an open problem, whether it is possible to define modelled risk variables $\bs Y_i$ such that 
\begin{itemize}
\item every modelled subrisk $\bs Y_i$ yields an appropriate risk capital model for the single subrisk $\bs X_i$ and
\item $\bs Y_{sum}=\sum \bs Y_i$ is an appropriate risk capital model for the overall risk $\bs X_{sum}$.
\end{itemize}
Moreover, our solution does only work for the multivariate normal distribution. A solution for other distributions relevant in practice is a topic for future research.  

\section{Declarations}
\textbf{Acknowledgements.} The experimental results have been generated using a Java program. We are grateful for the opportunity to run the program on the bwGriD cluster of the Hochschule Esslingen.\\
The work of the second author is supported by the DVfVW (Deutscher Verein f\"ur Versicherungswissenschaft) by a Modul 1 Forschungsprojekt with the title ``Das Parameterrisiko in Risikokapitalberechnungen f\"ur Versicherungsbest\"ande''.

\end{document}